\documentclass[12pt,a4paper]{article}
\usepackage{graphicx}
\begin{document}
\vspace*{5mm}
\begin{flushright}
 RRC \hspace*{13mm}\strut\\"Kurchatov Institute"\\
 IAE-6274/4 \hspace*{5mm}\strut
\end{flushright}
 \vspace*{15mm}
  \begin{center}
   \Large \bf $\pi\pi$  scattering at low energy:\\
    S-wave phase  shifts and  scattering  lengths.
  \end{center}
 \vspace*{1mm}
\begin{center}
  \large {V.N.Ma\u{\i}orov,  O.O.Patarakin}
\end{center}
\vspace*{1mm}
\begin{center}
\it  Russian  Research  Center "Kurchatov  Institute",\\
     Institute  of General  and  Nuclear  Physics,\\
     pl. Kurchatova 1, 123182 Moscow, Russian Federation\\
  E-mail: mvn@dnuc.polyn.kiae.su;  patarak@dnuc.polyn.kiae.su
 \end{center}
\begin{abstract}
On the basis  of the S-wave phase shifts  $\pi\pi$  scattering  behaviour from the threshold
up to dipion mass $m_{\pi\pi}=~1~GeV$, it is shown, that  the linear  correlation relates
the S-wave phase shifts: $\delta^0_0(s)=\eta\,\delta^2_0(s)$, where $\eta=-4.65\pm0.05$.
By using this correlation  at the solution of the Roy equations,
the accuracy of determination of S-wave   lengths $a^0_0$  and  $a^2_0$
is  considerably improved:
$a^0_0=(0.220\pm 0.006)\,m_{\pi}^{-1}$;  $a^2_0=(-0.0472\pm 0.0013)m_{\pi}^{-1}$.
The  obtained result unambiguously witnesses in favor of the standard ChPT with
large  quark condensate.
\end{abstract}
\vspace*{3mm}
\hspace*{26mm}  Perturbation  Theory\\[10mm]
\newpage
\section{Introduction}
The chiral symmetry of QCD Lagrangian is well known to be spontaneously broken down.
Two theories - Chiral Perturbation Theory (ChPT)  [1,2] and Generalised Chiral Perturbation
Theory (GChPT) [3] based on QCD - can describe the strong interactions at low energies.
The determinative factor in these theories is the existence of vacuum condensates violating
chiral symmetry. These theories having the same form of the effective  Lagrangian  differ
from each other by  value of  quark condensate and light quark masses.
The fact determining the choice of the version is that
the S-wave $\pi\pi$  scattering lengths $a^0_0$ and $a^2_0$ are very sensitive to the
parameters of the model and consequently are the key parameters for unambiguous
determination of the scenario of chiral symmetry violation. So, ChPT  predicts the value
$a^0_0$=0.220 and GChPT  $a^2_0=0.263$\footnote{The S-wave scattering lengths $a^0_0$
and $a^2_0$ are given in $m_{\pi}^{-1}$}.\\
During some time, despite large accumulated experimental material on scattering lengths,
this choice was difficult to be made. The matter is that the experiment $K_{e4}$ [4] gave
evidence in favor of GChPT, whereas most $\pi N\longrightarrow\pi\pi N$ experiments inclined
rather to ChPT.
The situation has changed recently as the latest experiment $K_{e4}$ E865 [5]
showed  evidence for ChPT, i.e. for the version  with strong quark condensate and small
masses of light quarks.
The present work shows that even without using the latest  $K_{e4}$ results,
based only on the information contained in experimental S-wave $\pi\pi$ scattering
data array, it is possible to improve considerably the accuracy of determination
of $a^0_0$ and $a^2_0$ and thus to choose the scenario of chiral symmetry violation
for certain.
%
\section{Roy  equations}
In our previous work [6] the  Roy  equations [7,8] were  used  to calculate the S-wave $\pi\pi$
scattering lengths. The experimental values  of the  S- and P-wave $\pi\pi$ phase shifts,
obtained  from an analysis of five charged channels, were presented  by  that time  in all
dipion mass region    from    the  threshold up to $m_{\pi\pi}=~1~GeV $ [9-22].
Based  on this fact we could realize the consecutive  solution technique of the  Roy
equations without using iterative procedures. Due to this  approach the problem of  convergence
of the  solutions was eliminated automatically  and the process of calculation of  scattering
lengths   $a^0_0 $ and $a^2_0 $ becomes absolutely clear.
%
 \begin{figure}[h]
   \includegraphics[width=\textwidth, height=4.5in]{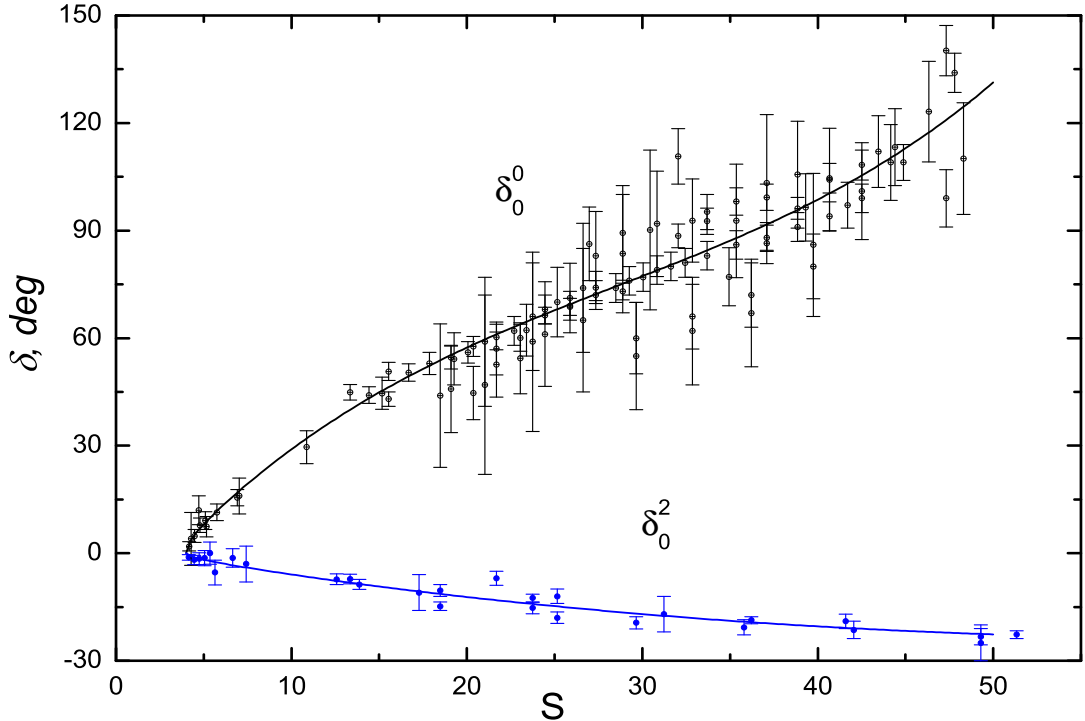}
   \caption{S-wave $\pi\pi$  phase  shifts. \quad  The solid curves  represent the result
    of  fitting  in  terms  of  expression (3).}
 \end{figure}
For the  case of the charged pions,   the  Roy equations are given by :
$$
  Ref^I_l(s)=\lambda^I_l(s)+\frac{1}{\pi}\,\int_4^{51}\Psi(x,s)\,dx+\varphi^I_l(s)
  \eqno  (1)
$$
Explicit expressions  for $\Psi (x,s\footnote{Here and below, s is the Mandelstam variable,
$s=m^2_{\pi\pi}/m^2_{\pi}$})$  are given  in [6].
The corrections  $\varphi^I_l (s)$,  estimating  the contributions from the  higher  waves
($l\ge2 $) and  from the large mass region were adopted from [23].
According to the theory [8]  subtractions  $ \lambda^I_0 (s) $  are defined as follow:
$$
  \lambda^0_0(s)=a^0_0+\frac{s-4}{12}\,(2a^0_0-5a^2_0); \qquad
  \lambda^2_0(s)=a^2_0-\frac{s-4}{24}\,(2a^0_0-5a^2_0) \eqno (2)
$$
The solution  of the  Roy  equations   (1)  comprised  some steps.
First, we performed fitting for each phase shifts $ \delta^I_l $ and obtained smooth  curves
adequately describing experimental data. In particular for the S-wave phase shifts
expansion  (3) was used:
$$
  \delta^I_0(s)=\frac{2}{\sqrt{s}}\,(C^I_1\,q+C^I_2\,q^3+C^I_3\,q^5+C^I_4\,q^7)
  \eqno (3)
$$
Where $q =\frac12\sqrt {s-4} $ -  is c.m. pion momentum  and $C^I_k$ - are free
parameters (I=0,2; $k=1\div4) $. Experimental values of phase shifts  and fitting curves (3)
are shown in Fig.1.
Then, the  resulting smooth curve  $ \delta^I_l (s) $ were input into the  Roy  equations (1)
to find  subtractions $\lambda^I_l(s)$. At the conclusive stage  we carried out fitting of the
subtractions  $\lambda^I_0(s)$  using  terms (2) and determined the S-wave $ \pi\pi$ lengths.
The resulting subtractions $\lambda^I_0(s)$  with fitting straight lines (2) are shown in Fig.2.
Note that the phase shifts  $ \delta^2_0 $ from  "electronic experiment"
[18] were not used. This question was considered  in [6] in detail.
%
 \begin{figure}[h]
   \includegraphics[width=\textwidth, height=4.5in]{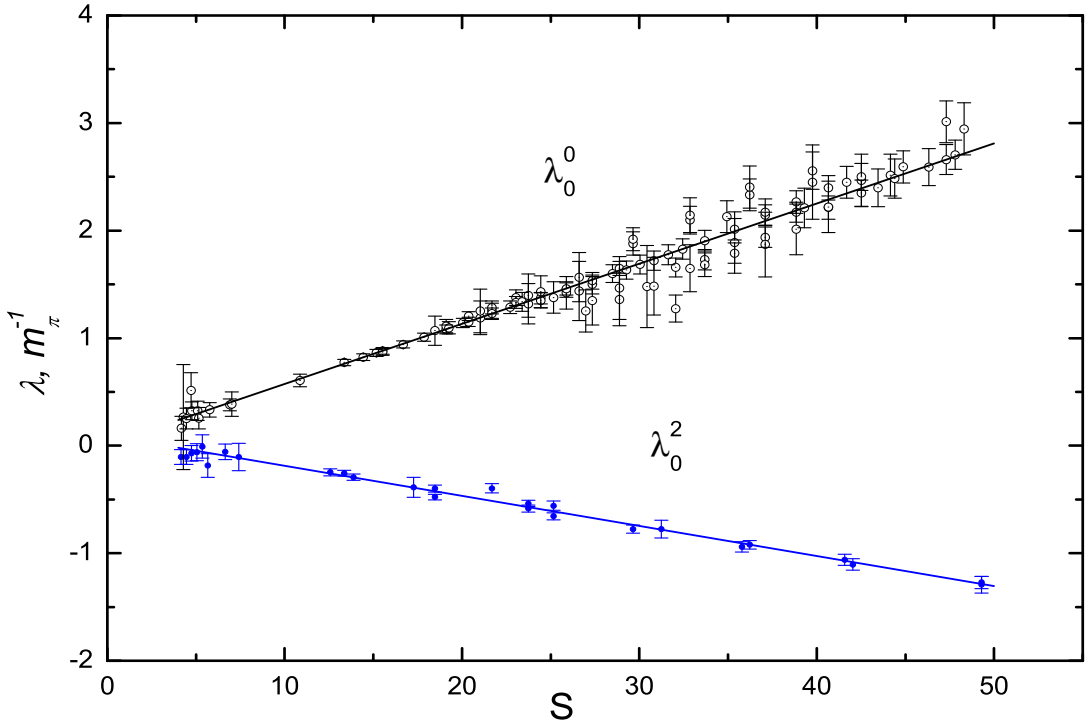}
   \caption{Subtractions $\lambda^0_0(s)$ and $\lambda^2_0(s)$.
   \ The straight  lines represent   the   result  of  fitting   in   terms   of
       expression (2).}
  \end{figure}
We  stress  that the linear relations  were obtained indeed when fitting
subtractions $ \lambda^I_0 (s) $. For us, it is a proof of that all the  calculation  steps
of the   solving the Roy equations and also all the preliminary work comprising  the fitting
phase shifts $ \delta^I_0(s) $   in terms (3) were carried out  correctly.
In our previous study [6]  we obtained:
$$
         a^0_0=0.240\pm0.023; \quad   a^2_0=-0.034\pm0.013 \eqno (4) \\
$$
with correlation coefficient   r=0.945.
We stress that in [6] we  made use of all available experimental  data, carried out the
consequent analysis, but resulting uncertainties of $a^0_0$ and $a^2_0$  were so large
that they did not allow to choose between  ChPT  and  GChPT.
%
\section{Correlation  between  $\delta^0_0(s)$ and $\delta^2_0(s)$}
The  analysis which has been carried out  showed that the general statistics contained in
the available data for the phase shifts is enough to improve significantly the accuracy
of determination of the S-wave scattering lengths and thus to choose correct ChPT version.
But the strong correlation between $a^0_0 $ and $a^2_0 $, which is
present in any method of the solution of the Roy equations,  in a latent form may
be, prevents one to make the above  and results in large uncertainties of the values
$a^0_0 $ and $a^2_0 $.

It should be noted, that the  correlation between $a^0_0 $ and $a^2_0 $ is present in both
scenarios ChPT and GChPT [24] due to causes of  fundamental nature.
The principal conclusion  can be made  that  using  model-independent analysis, based solely
on the Roy equations, it is impossible to increase the accuracy of definition of the S-wave
$\pi\pi$ lengths  $a^0_0 $ and $a^2_0 $ without removing
beforehand the correlation between them.
Therefore  we  need  some additional constraint on    $a^0_0 $ and $a^2_0 $.  And we shall
demonstrate that such constraint can be retrieved from  the available data.\\
Let us analyse our previous result (4).
As the correlation coefficient {\large r}  is very close to one,
  \begin{figure}[!t]
    \includegraphics[width=\textwidth, height=5in]{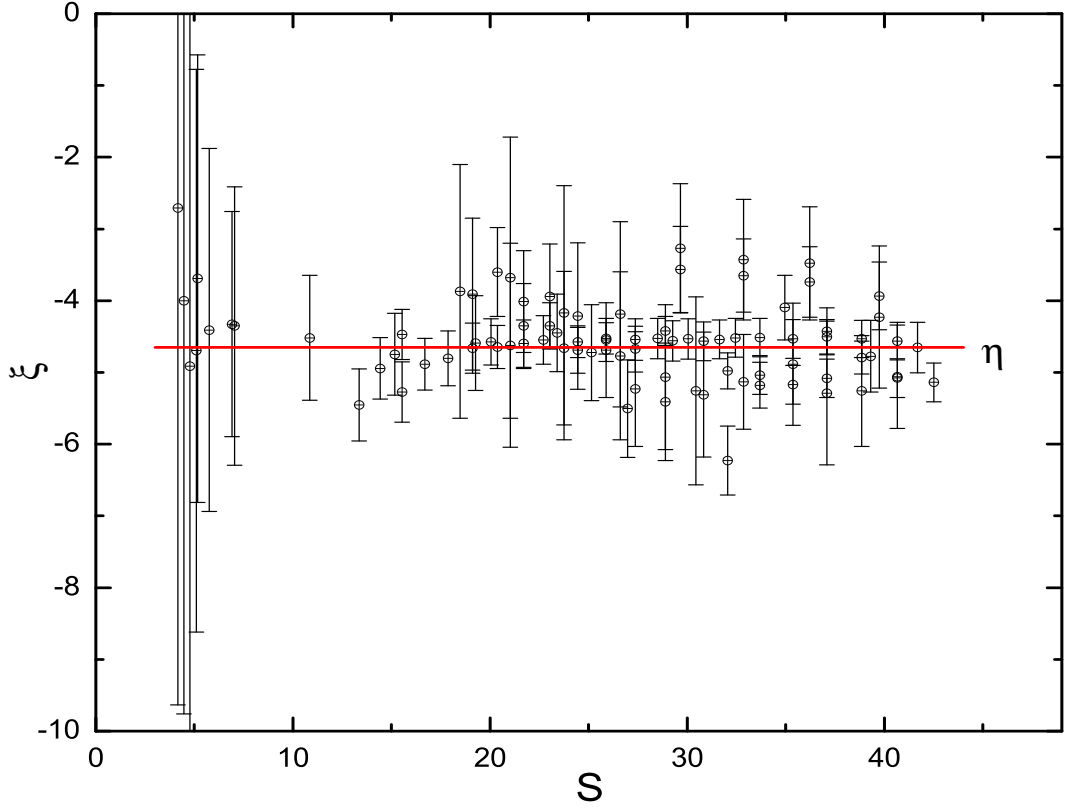}
    \caption{The ratio of the S-wave phase shifts $\xi(s)=\delta^0_0(s)/\delta^2_0(s).$
     The straight   line   represents   the   constant  $\eta=-4.65.$}
  \end{figure}
it implies the values $a^0_0 $ and $a^2_0 $ are  related by  linear dependence.
Then, as  $ \delta^I_0 (s) \propto a^I_0\,q $  near the threshold , the phase shifts
$ \delta^0_0 (s) $ and $ \delta^2_0 (s) $  must be related by linear dependence too.
In this way the simplest hypothesis, i.e. the hypothesis about
proportionality phase shifts into some area above  the threshold, must be verify.
For this the ratio  $\xi(s)=\delta^0_0(s)/\delta^2_0(s)$ was analysed for the available
experimental data.  As the phase shifts $\delta^0_0(s_i)$   and $\delta^2_0(s_j)$ were
measured mainly at different  energy values, the smoothed curve (Fig.1) representing
the fitting function (3)  was used for calculation of the phase shifts
$\delta^2_0$ at the points $s=s_i$, where the phase shifts $\delta^0_0$  were measured.
Thus, the ratio of the S-wave phase shifts was calculated as
$\xi (s_i) = \delta^0_0 (s_i)/\delta^2_0 (s=s_i)$.
The values  $ \xi (s_i) $ are given  in Fig.3. \\
The uncertainties  $ \sigma _ {\xi} $ were calculated  by the standard  rule of  propagation
of errors and finally they were defined both by errors of phase shifts $\delta^0_0(s_i)$
and  $\delta^2_0(s_j)$ .\\
The form of the dependence  $\xi (s)$  up to s=42, i.e. up to $m_{\pi\pi}$=900MeV,
makes one think that  $\xi (s)$ is  really a constant in this energy region.
It  witnesses  in  favor  of the    hypothesis being verified.
It  was  calculated  by fitting $\xi (s) \equiv\eta$--const  for the
interval S = $11\div42.5 $:
$\eta=-4.65\pm0.05$; \ $\chi^2$=75; \ $\langle\chi^2\rangle$=80.
  \begin{figure}[!t]
    \includegraphics[width=\textwidth, height=5in]{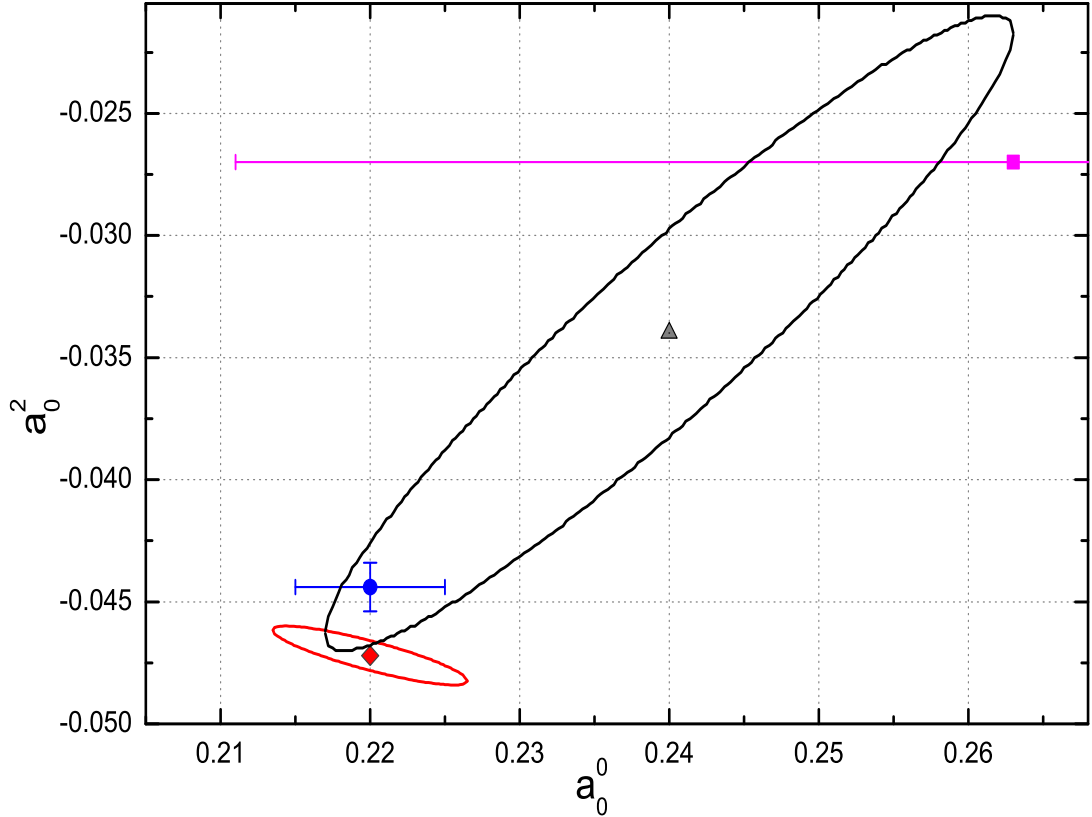}
    \caption{The  $\pi\pi$ scattering  lengths.  The ellipse with the centre as a triangle
      indicates our previous results [6]; the ellipse with the centre as a rhomb  shows  the
      result  of the present paper. The  cross  shows  the latest   ChPT  result [27];
      the square  shows  GChPT  result [3].}
 \end{figure}
So, the proposed   hypothesis is  proved by statistics and  consequently  we  can conclude
that  in the wide  enough  region  the ratio  of  the S-wave phase  shifts does not
depend on energy.
Thus   the new correlation    take  place  for  this  energy  region:

                   $$\delta^0_0(s)=\eta\,\delta^2_0(s)  \eqno  (5) $$
\noindent where $ \eta =-4.65\pm0.05 $.
As  stated above, from the fact of strong correlation of the scattering  lengths  follows
the linear dependence  of the phase  shifts near the threshold  and  such dependence
is found  in some  region above the threshold.
Based on these  facts we suppose that  the phase  shifts proportionality keeps constant
 for all  the energy region up to s=42.5.
So we extrapolate the relation (5) down to the threshold and assume that the factor
of  proportionality $\eta$  keeps  its value in close vicinity  to the threshold.
The  fitting $\xi (s) \equiv\eta$--const  for the interval s = $4\div42.5 $  gets  naturally
the same value of $\eta$,  because statistical weights  of the points near the threshold
are insignificant.
In general, vast uncertainties  of the values $ \xi (s) $ near the threshold (Fig.3)
are caused by the fact that the phase shifts  $ \delta^0_0 (s) $ and $\delta^2_0 (s)$
have large relative errors in that region.
And apparently  this situation    will  not change  in  the nearest  future.
If  the relation  (5)  is true  near the threshold, as we assume,  then  we obtain the new
constraint on the scattering   lengths:

                          $$a^0_0=\eta\,a^2_0   \eqno  (6)$$
This relation can be used to improve  the accuracy of determination  of the S-wave lengths.
\section{Calculation of scattering lengths  $a^0_0$  and $a^2_0$}
First, the estimation of the values $a^0_0$ and  $a^2_0$ was made without using the Roy
equations. Two  equations, the correlation (6) and the equation of the  "universal  curve" [25]
were solved together.
As a result we had:
$$
   a^0_0=0.20\pm0.02; \quad   a^2_0=-0.042\pm0.005  \eqno (7)
$$
It is necessary to note, that  in the work [26] in which the  "universal  curve" was introduced,
 attempt also was made to estimate independently the ratio $a^0_0/a^2_0$ for further
calculating the S-wave lengths $a^0_0$  and  $a^2_0$.\\
Then, the correlation (6) was used as a supplementary condition when solving the Roy equations.
And we obtained at the end:
$$
  a^0_0=0.220\pm0.006; \quad   a^2_0=-0.0472\pm0.0013  \eqno (8)
$$
The final result (8) together with our previous result (4) as well as the theoretical
predictions of ChPT and GChPT are presented in Fig.4.
%
\section{Conclusion}
1) On  the basis of the consequent statistical analysis  of the S-wave phase shifts $\pi\pi$
scattering  behaviour the new correlation (5) has been obtained which relates
the S-wave  phase shifts  $\delta^0_0(s)$ and  $\delta^2_0(s)$  from the threshold up
to   $m_{\pi\pi}=~900\,MeV$.  We believe that this correlation   has  value in itself for understanding
 the mechanism  of strong interactions.\\[1mm]
2)  The new constraint  (6) on the  S-wave scattering lengths $a^0_0$
 and $a^2_0$ has been computed  by  extrapolation  of the relation (5) to the threshold.\\[1mm]
3) The constraint (6) has been used as a supplementary condition when solving the Roy equations (1).
As the result the exact values of S-wave scattering lengths have been calculated:
$ a^0_0=0.220\pm0.006; \quad a^2_0=-0.0472\pm0.0013$.\\[1mm]
4)Comparison with the results of the work [27]: \ \ $a^0_0=0.220\pm0.005;$ \\
 $a^2_0=-0.0444\pm0.0010$,  where ChPT calculations  were supplemented with the phenomenological
 representations [25], shows that one more  argument is found for  the  standard
 ChPT [1,2]  with small light quark  masses and large quark condensate.\par \vspace{3mm}
This work was supported in part by the Russian Foundation for
Basic Research (project no. 00-02-17852).

 \end{document}